\documentclass[prl,twocolumn,preprintnumbers,superscriptaddress,showpacs]{revtex4-1}

\usepackage{graphicx}
\usepackage{bm}
\usepackage{amsmath}
\usepackage{braket}
\usepackage{diagbox}
\usepackage{booktabs}
\usepackage{amssymb}
\usepackage{slashed}
\usepackage[colorlinks,pdfstartview=FitH]{hyperref}
\hypersetup{linkcolor=blue,citecolor=blue,filecolor=black,urlcolor=blue}

\def \a {\alpha}
\def \b {\beta}
\def \c {\chi}
\def \d {\delta}

\def \g {\gamma}
\def \h {\eta}

\def \p {\pi}
\def \ps {\psi}

\def \r {\rho}
\def \l {\lambda}
\def \m {\mu}
\def \n {\nu}
\def \s {\sigma}

\def \th {\theta}
\def \x {\xi}

\def \f {\phi}
\def \vf {\varphi}

\def \D {\Delta}
\def \G {\Gamma}
\def \O {\Omega}
\def \P {\Pi}
\def \L {\Lambda}

\def \<{\langle}
\def \>{\rangle}
\def \+{\dagger}
\def \({\left(}
\def \){\right)}
\def \[{\left[}
\def \]{\right]}

\def \vp {\bf{p}}

\def \jps {{J/\ps}}
\def \hB {\hat{B}}
\def \bB {{\bf B}}

\def \br {{\bf r}}
\def \bK {{\bf K}}
\def \bP {{\bf P}}
\def \bq {{\bf q}}
\def \bk {{\bf k}}
\def \Tr {{\text{Tr}}}
\def \bmu {{\boldsymbol\mu}}

\def \fnr {\phi_{\text{NR}}}
\def \cnr {\chi_{\text{NR}}}

\begin{document}

\author{Guowei Yan}
\affiliation{School of Physics and Astronomy, Sun Yat-Sen University, Zhuhai 519082, China}
\author{Shu~Lin}
\affiliation{School of Physics and Astronomy, Sun Yat-Sen University, Zhuhai 519082, China}
\affiliation{Guangdong Provincial Key Laboratory of Quantum Metrology and Sensing, Sun Yat-Sen University, Zhuhai 519082, China}

\title{Distorted quarkonia and spin alignment}
\date{\today}


\begin{abstract}
	It is well-known that atoms can change their shape when subject to external electromagnetic fields. Analogous phenomenon is expected for particles, which are much smaller in size with the corresponding shape change much harder to observe. We point out that shape change of particles can be accessed by measurements of their spin alignment. Motivated by recent measurements of spin alignment in relativistic heavy ion collisions, we consider quarkonia in electromagnetic fields as an example. We show that the quarkonia spin alignment receives both spin contribution from spin states mixing and orbital contribution from shape change. By a proper choice of quantization axis, it is possible to switch off the spin contribution, leaving only the orbital contribution. This makes spin alignment measurement a valuable probe of quarkonium structure.
\end{abstract}

\maketitle

\section{Introduction}\label{sec1}

The structure of a composite particle is one fundamental property. Traditionally, it is measured through scattering with other charged particles, in which the exchanged photon probes the distributions of electric charge and magnetic moment \cite{Hofstadter:1958psf}. When external electromagnetic (EM) fields reach QCD scale, the particle can also respond in shape change and spin polarization like an atom. Such strong EM fields are indeed produced in relativistic heavy ion collisions (HIC) \cite{Skokov:2009qp,Deng:2012pc}.

Recently, multiple measurements of spin polarization for $\L$ hyperon \cite{STAR:2017ckg} and spin alignment for $\f$ and $\jps$ in HIC \cite{STAR:2022fan,ALICE:2019aid,ALICE:2022dyy} have been reported, following early predictions based on spin-orbit coupling \cite{Liang:2004ph,Liang:2004xn}. Indeed, this picture nicely explains global spin polarization in HIC with a vortical quark-gluon plasma (QGP) \cite{Betz:2007kg,Becattini:2007sr,Becattini:2013vja,Pang:2016igs}. The mechanism of spin alignment is more dependent on particle species: $\f$ are produced at late stage thus is more susceptible to properties of QGP \cite{Sheng:2022wsy,Sheng:2023urn,Kumar:2023ghs,Li:2022vmb,Wagner:2022gza}; $\jps$ are produced at early stage and thus can be affected by multiple sources in the evolution of HIC \cite{Zhao:2023plc,DeMoura:2023jzz,Yang:2024ejk,Sheng:2024kgg,Zhao:2024ipr,Chen:2025mrf,Liang:2025hxw,Ahmed:2025bwi,Grossi:2024pyh,Sahoo:2025bkx,Dey:2025ail}. In particular, strong EM fields produced at early stage of HIC are expected to be a significant source for spin alignment. Indeed, they are found to be responsible for direct flow of heavy mesons \cite{STAR:2019clv,ALICE:2019sgg,Shen:2025unr}. The EM fields can contribute to spin alignment either by introducing imbalance in spin states or by changing particle shape, corresponding to spin and orbital contributions respectively. The latter has been ignored so far. The purpose of this paper is to show that orbital contribution is not only allowed, but also can be separated, offering a valuable probe of particle structure.

\begin{figure}[h]
	\centering
	\includegraphics[width=0.95\linewidth]{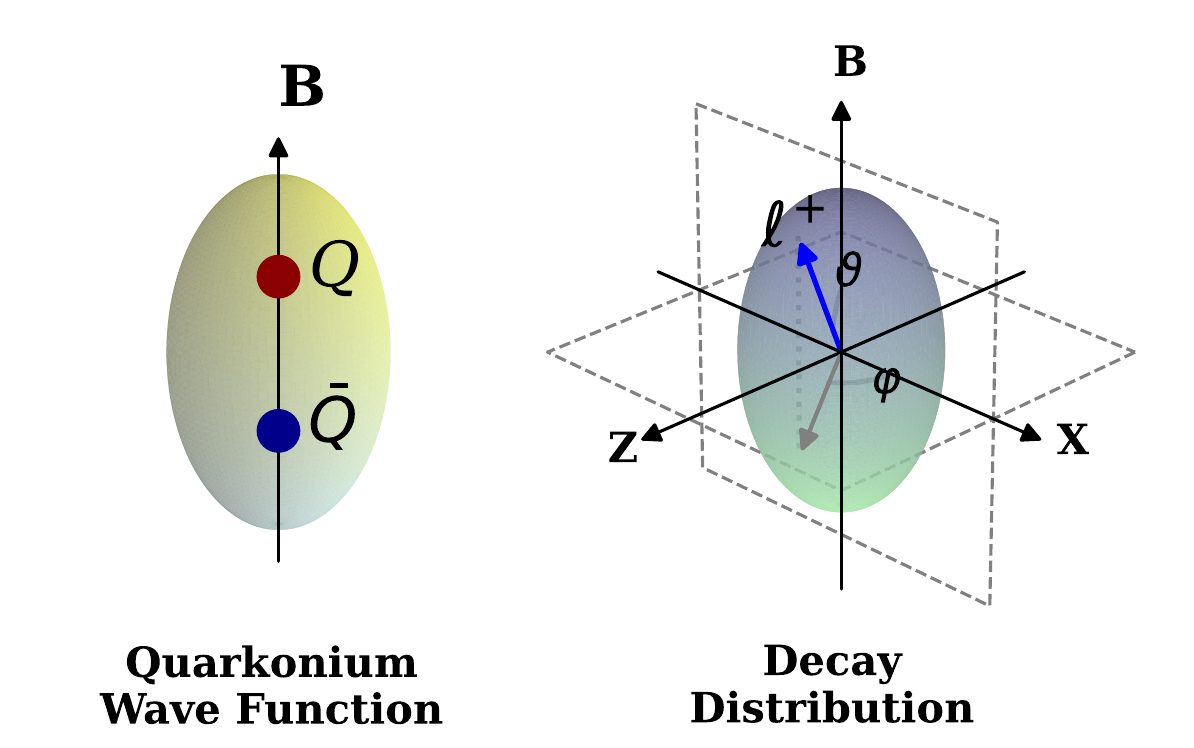}
	\caption{Magnetic field distorts wave function of quarkonium, leading to anisotropic distribution of its decay product. Magnetic field direction is chosen as quantization axis.}\label{fig:illustration}
\end{figure}

Motivated by measurements in HIC, we study EM contribution to spin alignment of $\jps$-like quarkonia, i.e. S-wave spin triplet state. On one hand, the magnetic field induces mixing between spin triplet and singlet states, leading to imbalance among the triplet states; on the other hand, the EM fields can distort the quarkonia spatial wave function, as a manifestation of shape change. We will see the original S-wave state is changed into a superposition of S-wave and D-wave states. Since quarkonia spin alignment is measured through angular distribution of daughter leptons, the resulting anisotropic wave function naturally gives rise to orbital contributions to spin alignment, as illustrated in Fig.~\ref{fig:illustration}. We shall show in this work that the spin alignment can be a much more sensitive probe of quarkonium potential than the conventional method.



\section{Anisotropic photon and spin alignment}

Consider one quarkonium state annihilating into lepton pair, the differential rate is given by
\begin{align}\label{diff_rate}
&\frac{d\G}{d^4P}=e^4\int_{{\bf p}_1,{\bf p}_2}\hspace{-5mm}(2\p)^4\d^4(p_1+p_2-P)|\<l\bar{l}|J^\a D_{\a\m} J^\m|Q\bar{Q}\>|^2\nonumber\\
&=\frac{e^4}{(P^2)^2}\int_{{\bf p}_1,{\bf p}_2}\hspace{-5mm}(2\p)^4\d^4(p_1+p_2-P) J^\m|Q\bar{Q}\>\<Q\bar{Q}|J^\n l_{\m\n},
\end{align}
where $p_{1,2}$ are momenta of lepton and anti-lepton respectively and $\int_{{\bf p}_1,{\bf p}_2}=\int\frac{d^3p_1}{2p_1(2\p)^3}\frac{d^3p_2}{2p_2(2\p)^3}$. The quarkonium and dilepton states are connected by intermediate photon states with $D_{\a\m}=\frac{-i\h_{\a\m}}{P^2}$ being the photon propagator in Feynman gauge and $l_{\m\n}=4\[p_{1\m}p_{2\n}+p_{1\n}p_{2\m}-(p_1\cdot p_2)\h_{\m\n}\]$ is the final state lepton tensor \cite{Bellac:2011kqa}. The derivation of \eqref{diff_rate} assumes a specific initial state and $|Q\bar{Q}\>\<Q\bar{Q}|$ is the corresponding density matrix. It is easily generalized to more general initial state by the replacement of the density matrix: $|Q\bar{Q}\>\<Q\bar{Q}|\to\r_{AB}|Q\bar{Q}\>_A\,{}_B\<Q\bar{Q}|$ with $A,\,B$ labeling the spin triplet states and $\r_{AB}$ denoting the spin density matrix. Let's define
\begin{align}\label{Pi_amp}
\P^{\m\n}=\r_{AB}J^\m|Q\bar{Q}\>_A\,{}_B\<Q\bar{Q}|J^\n.
\end{align}
as the photon self-energy for the quarkonium state. It is constrained by Ward identity as $p_\m\P^{\m\n}=0$. In the rest frame where $p_\m=(M,0)$, we have $\P^{00}=\P^{0i}=0$. The remaining rotational invariance dictates $\P^{ij}\propto\d^{ij}$. In the presence of magnetic field that breaks rotational symmetry, we can parameterize the self-energy as
\begin{align}\label{Pi_TL}
\P^{ij}=(\d^{ij}-\hB^i\hB^j)\P_T+\hB^i\hB^j\P_L.
\end{align}
Performing the phase space integration, we obtain the following angular distribution
\begin{align}\label{angular}
&\frac{d\G}{d\cos\th}=\frac{e^4}{16\p M^2}\big[\P_T(3-\cos^2\th_{Bl})+\P_L(1+\cos^2\th_{Bl})\nonumber\\
&+\cos^2\th\((\P_T-\P_L)(2\cos^2\th_{Bl}-\sin^2\th_{Bl})\)\big],
\end{align}
where $\th$ is the angle between lepton momentum to a quantization axis $\hat{l}$ and $\th_{Bl}$ is the angle between $\hat{B}$ and $\hat{l}$.
In the absence of magnetic field, $\P_L=\P_T$, the angular distribution is isotropic giving vanishing spin alignment. When $\P_L\ne\P_T$, a nonvanishing spin alignment follows as
\begin{align}\label{alignment}
\l_\th=\frac{(\P_T-\P_L)(3\cos^2\th_{Bl}-1)}{\P_T(3-\cos^2\th_{Bl})+\P_L(1+\cos^2\th_{Bl})}.
\end{align}

\section{Photon self-energy for quarkonium state}

To evaluate the photon self-energy, we need to calculate $J^i|Q\bar{Q}\>_A$, which is the amplitude for current $J^i$ to annihilate a quarkonium in spin state $A$. We shall express the quarkonium state using quantum mechanical language \cite{Peskin:1995ev}
\begin{align}\label{state_decomp}
|Q\bar{Q}\>_A=\sqrt{2M}\int\frac{d^3q}{(2\p)^3}\ps(\bq)\frac{1}{2m}|\bq,-\bq\>_A,
\end{align}
with $\ps(\bq)$ being the spatial wave function and $|\bq,-\bq\>_A$ denoting a plane wave in spin state $A$. Note that in this description, quarkonium is written in terms of particle basis, in which particle and anti-particle decouple \cite{Bodwin:1994jh}. We shall express the current also in the particle basis. For our purpose, it is sufficient to derive the current in free theory, for which the quark field in Dirac representation reads \cite{Itzykson:1980rh}
\begin{align}
\ps(x)=\int\frac{d^3p}{(2\p)^3\sqrt{2E_p}}\sum_s\(a_{\vp}^s u_s(p)e^{-ip\cdot x}+b^{s\dagger}_{\vp} v_s(p)e^{ip\cdot x}\),\nonumber
\end{align}
with the spinors given by
\begin{align}
u_s(p)=\begin{pmatrix}
\frac{m+E_p}{\sqrt{2m(m+E_p)}}\x_+^s\\
\frac{{\bf p}\cdot{\boldsymbol{\s}}}{\sqrt{2m(m+E_p)}}\x_+^s
\end{pmatrix},
v_s(p)=\begin{pmatrix}
\frac{{\bf p}\cdot{\boldsymbol{\s}}}{\sqrt{2m(m+E_p)}}\x_-^s\\
\frac{m+E_p}{\sqrt{2m(m+E_p)}}\x_-^s
\end{pmatrix}.
\end{align}
$\x_+^s$ and $\x_-^s$ are spinor doublets in spin $s$ state for quark and anti-quark respectively. We shall suppress the spin indices $s$ below. We can eliminate the small components of $u$ and $v$ by the Foldy-Wouthuysen (FW) transformation \cite{Foldy:1949wa}
%
\begin{align}
u'=U_u u,\quad v'=U_v v.
\end{align}
The corresponding FW transformations are given by $U_u=U(\vp)$ and $U_v=U(-\vp)$ with $U({\bf p})=\cos\th+\b\frac{{\boldsymbol{\a}}\cdot{\bf p}}{|p|}\sin\th$. $\th$ is fixed by $\cos2\th=\frac{m}{E_p}$ and $\sin2\th=\frac{|p|}{
	E_p}$. It leads to transformed spinors with large component only
\begin{align}
u'=\begin{pmatrix}
\sqrt{\frac{E_p}{m}}\x_+\\
0
\end{pmatrix}\equiv
\begin{pmatrix}
&\fnr\\
&0
\end{pmatrix}\\
v'=\begin{pmatrix}
&0\\
&\sqrt{\frac{E_p}{m}}\x_-
\end{pmatrix}\equiv
\begin{pmatrix}
&0\\
&\cnr
\end{pmatrix}.
\end{align}
We have identified the resulting nonvanishing components as the non-relativistic wave functions in particle basis. To derive the current vertex in terms of the non-relativistic fields, we calculate the annihilation amplitude $J^i|Q\bar{Q}\>$ in quarkonium rest frame
\begin{align}
&J^i|Q\bar{Q}\>=\bar{v}(p')\g^i u(p)\nonumber\\
&=\cnr^\dagger\(\s^i-\frac{1}{E_p(m+E_p)}p^i\vp\cdot{\boldsymbol{\s}}\)\fnr,
\end{align}
where $p^\m=(E_p,\vp)$, $p'^\m=(E_p,-\vp)$ have been used. This gives the vertex $\D^i(p)=\s^i-\frac{1}{E_p(m+E_p)}p^i\vp\cdot{\boldsymbol{\s}}$. In the supplementary information, we show that the vertex, when expanded to $O(\frac{p^2}{m^2})$, agrees with the non-relativistic QCD (NRQCD) counterpart \cite{supplement}.

To proceed, we denote the annihilation amplitude $J^i|\bq,-\bq\>_A=\Tr[\D(q)^i(\cnr^\dagger\fnr)_A]$, with $A$ labeling the spin of the triplet state formed by the non-relativistic fields. Plugging this into \eqref{Pi_amp} and using \eqref{state_decomp}, we obtain upon dropping irrelevant constant to find
\begin{align}\label{probability}
&J^i|Q\bar{Q}\>_A\propto\int_{{\bq}}\ps(\bq)\text{Tr}[\D^i(q)(\x_+\x_-^\dagger)_A]\nonumber\\
&\to\int_{{\bq}}\ps(\bq)\text{Tr}[\D^i(q)\frac{1}{\sqrt{2}}{\bf n}^*_A\cdot {\boldsymbol{\s}}],
\end{align}
In the second line, the following replacements has been used \cite{Peskin:1995ev}
\begin{align}
&|11\>:\,\x_-\x_+^\dagger\to\frac{1}{\sqrt{2}} {\bf n}_+^*\cdot\boldsymbol{\s},\nonumber\\
&|1-1\>:\,\x_-\x_+^\dagger\to\frac{1}{\sqrt{2}} {\bf n}_-^*\cdot\boldsymbol{\s},\nonumber\\
&|10\>:\,\x_-\x_+^\dagger\to\frac{1}{\sqrt{2}} {\bf n}_0^*\cdot\boldsymbol{\s},
\end{align}
with ${\bf n}_\pm=\frac{1}{\sqrt{2}}(1,\pm i,0)$ and ${\bf n}_0=(0,0,1)$ when taking the quantization axis along $\hat{z}$.
%
\eqref{probability} leads to the following self-energy
\begin{align}\label{Pi_rep}
\P^{ij}\propto\r_{AB}\int_{\bq,\bk}\Tr[\D^i{\bf n}_A^*\cdot{\boldsymbol{\s}}\ps(\bq)]\Tr[\D^j{\bf n}_B\cdot{\boldsymbol{\s}}\ps^*(\bk)].
\end{align}
Anisotropic self-energy can arise either from distortion of quarkonium wave function $\ps$ or modification of spin density matrix $\r_{AB}$, which we discuss below.

\section{Quarkonium distortion and spin states mixing}%

The quarkonium Hamiltonian in background magnetic field can be expressed as \cite{Alford:2013jva}
\begin{align}\label{Hamiltonian}
&H=-\frac{\nabla^2}{m}+V(\br)+2m+\frac{\bK^2}{4m}-\frac{q}{2m}(\bK\times\bB)\cdot\br\nonumber\\
&+\frac{q^2}{4m}(\bB\times\br)^2-\bmu\cdot\bB,
\end{align}
where $\br$, $\bK$ and $\bmu$ are the relative coordinate, pseudomomentum and magnetic moment of the quarkonium. $\bK$ is conserved in the presence of Lorentz force and is related to the kinetic momentum $\bP$ by the operator relation $\bP=\bK-q\bB\times\br$. The first three terms correspond to quarkonium in the absence of magnetic field. The middle two terms are due to center of mass motion. The last two terms correspond to diamagnetic and Zeeman interactions respectively. We shall treat the last four terms as perturbations, which modify either spatial wave function or spin density matrix. In quarkonium rest frame, we can simply take $\bK=0$, which leads to a potential invariant under $\br\to-\br$. It follows from symmetry that $\<\bP\>=\bK-q\bB\times\<\br\>=0$ \footnote{Explicit confirmation of the choice can be found for the harmonic oscillator potential in \cite{Alford:2013jva}}.

Now we give a quantitative discussion on the effect of the remaining perturbations. We will first determine the photon self-energy $\P^{ij}$, and then apply \eqref{angular} to obtain the spin alignment. We start with the distortion effect from the diamagnetic interaction. Note that $\P^{ij}$ in \eqref{Pi_TL} is covariant and independent of the quantization axis. Thus we can point the magnetic field along $\hat{z}$ temporarily for the determination of $\P^{ij}$. In this case, the diamagnetic interaction reads $\D H=\frac{q^2}{4m}B^2r^2\sin^2\th$. The first order perturbation leads to the selection rules $\D l=0,2$ and $\D m=0$ for the eigenstates $|nlm\>$ of the unperturbed Hamiltonian, with $n$ and $l,m$ being principle and angular momentum quantum numbers respectively. The perturbation gives rise to D-wave and S-wave corrections to the unperturbed state as
\begin{align}
&\ps_D({\bf q})=\sum_n\frac{\<n20|\D H|100\>}{E_{n20}-E_{100}}\<\bq|n20\>,\nonumber\\
&\ps_S({\bf q})=\sum_n\frac{\<n00|\D H|100\>}{E_{n00}-E_{100}}\<\bq|n00\>,\label{Diaexpand}
\end{align}
where $E_{n20}$ and $E_{100}$ are energies of $|n20\>$ and $|100\>$ respectively. Since the correction enters the wave function, we may simply use the unperturbed spin density matrix $\r_{AB}=\frac{1}{3}\d_{AB}$. Using $\d_{AB}n_A^{*k}n_B^l=\d^{kl}$, we have from \eqref{Pi_rep}
\begin{align}\label{Pi_diamagnetic}
&\P^{ij}
\propto\d^{kl}\int_{\bq,\bk}\(\d^{ik}+\a_q\hat{q}^i\hat{q}^k\)\(\d^{jl}+\a_k\hat{k}^j\hat{k}^l\)\ps(\bq)\ps^*(\bk)\nonumber\\
&=\int_{\bq,\bk}\(\d^{ij}+\a_q\hat{q}^i\hat{q}^j+\a_k\hat{k}^i\hat{k}^j+\a_q\a_k\hat{q}^i\hat{k}^j\hat{q}\cdot\hat{k}\)\times\nonumber\\
&(\ps_0(\bq)+\ps_S(\bq)+\ps_D(\bq))(\ps_0(\bk)+\ps_S(\bk)+\ps_D(\bk)),
\end{align}
with $\a_q=-\frac{q^2}{E_q(m+E_q)}$ and $\ps_0(\bq)=\<\bq|100\>$ is the unperturbed wave function. Writing $\ps_0(\bq)=Y_{0}^0(\O_q)R_0(q)$, $\ps_S(\bq)=Y_{0}^0(\O_q)R_S(q)$ and $\ps_D(\bq)=Y_{2}^0(\O_q)R_D(q)$, with spherical harmonics $Y_l^m(\O_q)$, we can perform the integration of angular variable $\O_q$ using the following relations:
\begin{align}
&\int d\O_qY_{0}^0(\O_q)\hat{q}^i\hat{q}^j=\frac{1}{3}\d^{ij}(4\p)^{1/2},\nonumber\\
&\int d\O_qY_{2}^0(\O_q)\hat{q}^i\hat{q}^j=\(\hB^i\hB^j-\frac{1}{3}\d^{ij}\)\(\frac{4\p}{5}\)^{1/2},
\end{align}
We then obtain from \eqref{Pi_diamagnetic} up to first order in perturbation
\begin{align}\label{Pi_exp}
&\P^{ij}\propto\int_{k}\bigg[\d^{ij}\(1+\frac{\a_k}{3}\)(R_0(k)+2R_S(k))\nonumber\\
&+\frac{2}{\sqrt{5}}\a_k\(\hB^i\hB^j-\frac{1}{3}\d^{ij}\)R_D(k)\bigg],
\end{align}
with $\int_k\equiv\int k^2dk$. In arriving at the above, symmetry under $\bq\leftrightarrow\bk$ has been used and an overall factor $\int_q(1+\frac{\a_q}{3})R_0(q)$ has been dropped. We stress that $\P_{T/L}$ extracted from \eqref{Pi_exp} using \eqref{Pi_TL} are independent of our choice of $\hat{B}$ direction. Now we can plug the resulting $\P_{T/L}$ into \eqref{alignment}, which is applicable for arbitrary $\hat{B}$ and $\hat{l}$, to obtain the following spin alignment up to $O(B^2)$
\begin{align}\label{alignment_diamagnetic}
\l_\th\simeq -\frac{2}{\sqrt{5}}\frac{\int dk k^2 \a_k R_D(k)}{\int dk k^2(1+\frac{\a_k}{3}) R_0(k)}\frac{3\cos^2\th_{Bl}-1}{4}.
\end{align}
Note that the correction to S-wave doesn't contribute to leading order in the spin alignment. $\th_{Bl}=0$ corresponds to choosing the quantization axis perpendicular to the event plane, for which significant spin alignment is observed \cite{ALICE:2022dyy}\footnote{Other choices of quantization axis have also been used \cite{ALICE:2020iev,Massacrier:2024fgx}, yet with no significant signal of spin alignment.}.

To obtain the radial part of the wave function in \eqref{alignment_diamagnetic}, we first solve the Schr\"odinger equation for unperturbed eigenstates 
\begin{equation}
-\frac{1}{2\mu}\frac{d^2U_{nl}(r)}{dr^2}+\left[\frac{l(l+1)}{2\mu r^2}+V(r)\right]U_{nl}(r)=E_nU_{nl}(r),
\end{equation}
where the reduced radial wave function $U_{nl}(r)$ is defined as $U_{nl}(r)= rR_{nl}(r)$, and $\mu= m/2$ is the reduced mass of quarkonium. The potential $V$ is chosen to be the Cornell form 
\begin{equation}
V(r)=-\frac{\kappa}{r}+\frac{r}{a^2}+V_0.
\end{equation}
The parameters for $\jps$ are chosen as $\kappa=0.61,a=2.38\text{GeV}^{-1}$, $V_0=0$ and $m=1.84\text{GeV}$  \cite{Eichten:1978tg,Eichten:1979ms,Eichten:1995ch,Eichten:1994gt}.
The reduced radial wave functions $U_{nl}(r)$ along with the binding energy $E_n$ for each level are solved numerically following the method in \cite{Lucha:1998xc}. The transition matrix element in \eqref{Diaexpand} can be calculated in coordinate space as
\begin{align}
&\bra{n20}\Delta H\ket{100}\nonumber\\
&=\frac{q^2B^2}{4m}\int_0^\infty dr\, U_{n2}(r)r^2U_{10}(r)\nonumber\\
&\times \int_0^{2\pi} d\phi\int_0^\pi d\theta\sin\theta\,Y_2^0(\theta,\phi)\sin^2\theta\, Y_0^0(\theta,\phi).
\end{align}
%
In performing the summation in \eqref{Diaexpand} for $\ps_D$, we find the transition coefficient $c_n=\frac{\<n20|\D H|100\>}{E_{n20}-E_{100}}$ converges quickly,
as shown in TABLE \ref{TABLEFORDIA}. 
Accurate $\psi_D$ is obtained by summing over energy levels up to $n=15$. The radial part of the wave function in momentum space in \eqref{alignment_diamagnetic} can be computed with the reduced radial wave function in position space with
\begin{align}
{R}_{nl}(q) = \mathcal{N}_{nl} \int_0^\infty dr \,r U_{nl}(r) j_{l}(qr),
\end{align}
where $j_l(qr)$ is the spherical Bessel function of the first kind. $\mathcal{N}_{nl}$ is the nomalization factor. The angular part of the wave function in momentum space is the same as the case in position space \cite{Kirchbach:2022ptv}.
%
\begin{table}
	\setlength{\tabcolsep}{10pt}
	\begin{tabular}{|c|cccc|}
		\hline
		n & 3& 4& 5&6\\
		\hline
		$c_n$& 
		0.109& 0.0206& 0.00673&0.00341\\
		\hline
		n&7&8&9&10\\
		\hline
		$c_n$&0.00194&0.00112&0.000679&0.000461\\
		\hline
	\end{tabular}\label{TABLEFORDIA}
\caption{Transition coefficient $c_n=\frac{\<n20|\D H|100\>}{E_{n20}-E_{100}}$ for diamagnetic interaction in units of $(e|\bB|)^2/\text{GeV}^4$ for the first few transitions. Parameters for $J/\psi$ has been used for the quarkonium. }
\end{table}

Other than the diamagnetic interaction, the distortion effect can also occur from motion of quarkonia with respect to the magnetic field. In its own frame, the quarkounium sees an electric field from boosted magnetic field as ${\bf E}=\g{\bf v}\times \bB$ with ${\bf v}$ being the quarkonium velocity and $\g=(1-v^2)^{-1/2}$. The latter couples to the electric dipole of quarkonium and introduces the following term to the Hamiltonian
\begin{align}
\D H=q{\bf E}\cdot \br.
\end{align}
This leads to the motional Stark effect on the quarkonia. The selection rule for first order perturbation is $\D l=1$ and $\D m=0$, giving rise to a P-wave correction to the wave function. The parity odd correction doesn't contribute to self-energy as the corresponding integral in \eqref{Pi_rep} simply vanishes by parity. At second order, the perturbation gives rise to D-wave correction as
\begin{align}\label{Stark}
&\ps^E_D({\bf q})=\sum_{n,m}\frac{\<n20|\D H|m10\>\<m10|\D H|100\>}{(E_{n20}-E_{m10})(E_{m10}-E_{100})}\<\bq|n20\>-\nonumber\\
&\sum_{m}\frac{\<100|\D H|100\>\<m10|\D H|100\>}{(E_{m10}-E_{100})^2}\<\bq|m10\>,
\end{align}
with the superscript $E$ indicating its origin from the Stark effect. The second term vanishes as $\<100|\D H|100\>=0$ by parity. There is also a similar correction to S-wave $\ps_S^E(\bq)$. With Stark effect, the following decomposition of self-energy should be used
\begin{align}
\P^{ij}=(\d^{ij}-\hat{E}^i\hat{E}^j)\P_T+\hat{E}^i\hat{E}^j\P_L.
\end{align}
Using covariance and quantization axis independence of the self-energy, we point $\hat{E}$ along $z$-axis in calculating $\P^{ij}$. 
We again denote the wave corrections as $\ps^E_D(\bq)=Y_{2}^0(\O_q)R^E_D(q)$ and $\ps^E_S(\bq)=Y_{2}^0(\O_q)R^E_S(q)$, with superscripts $E$ indicating its origin from the Stark effect. Similar derivation of self-energy as in the diamagnetic interaction case applies. We end up with
\begin{align}\label{Pi_exp_E}
&\P^{ij}\propto\int_{k}\bigg[\d^{ij}\(1+\frac{\a_k}{3}\)(R_0(k)+2R^E_S(k))\nonumber\\
&+\frac{2}{\sqrt{5}}\a_k\(\hat{E}^i\hat{E}^j-\frac{1}{3}\d^{ij}\)R^E_D(k)\bigg],
\end{align}
from which $\P_L$ and $\P_T$ are easily extracted. Note that \eqref{alignment} still applies but with $\th_{Bl}\to\th_{El}$, thus we arrive at the following spin alignment from the Stark effect.
\begin{align}\label{alignment_Stark}
\l_\th\simeq -\frac{2}{\sqrt{5}}\frac{\int dk k^2 \a_k R^E_D(k)}{\int dk k^2(1+\frac{\a_k}{3}) R_0(k)}\frac{3\cos^2\th_{El}-1}{4}.
\end{align}
In case of heavy ion collision experiments, we have $\hat{l}\parallel\hat{B}$ so $\th_{El}=\frac{\p}{2}$.

\begin{figure}
	\centering
	\includegraphics[width=1.0\linewidth]{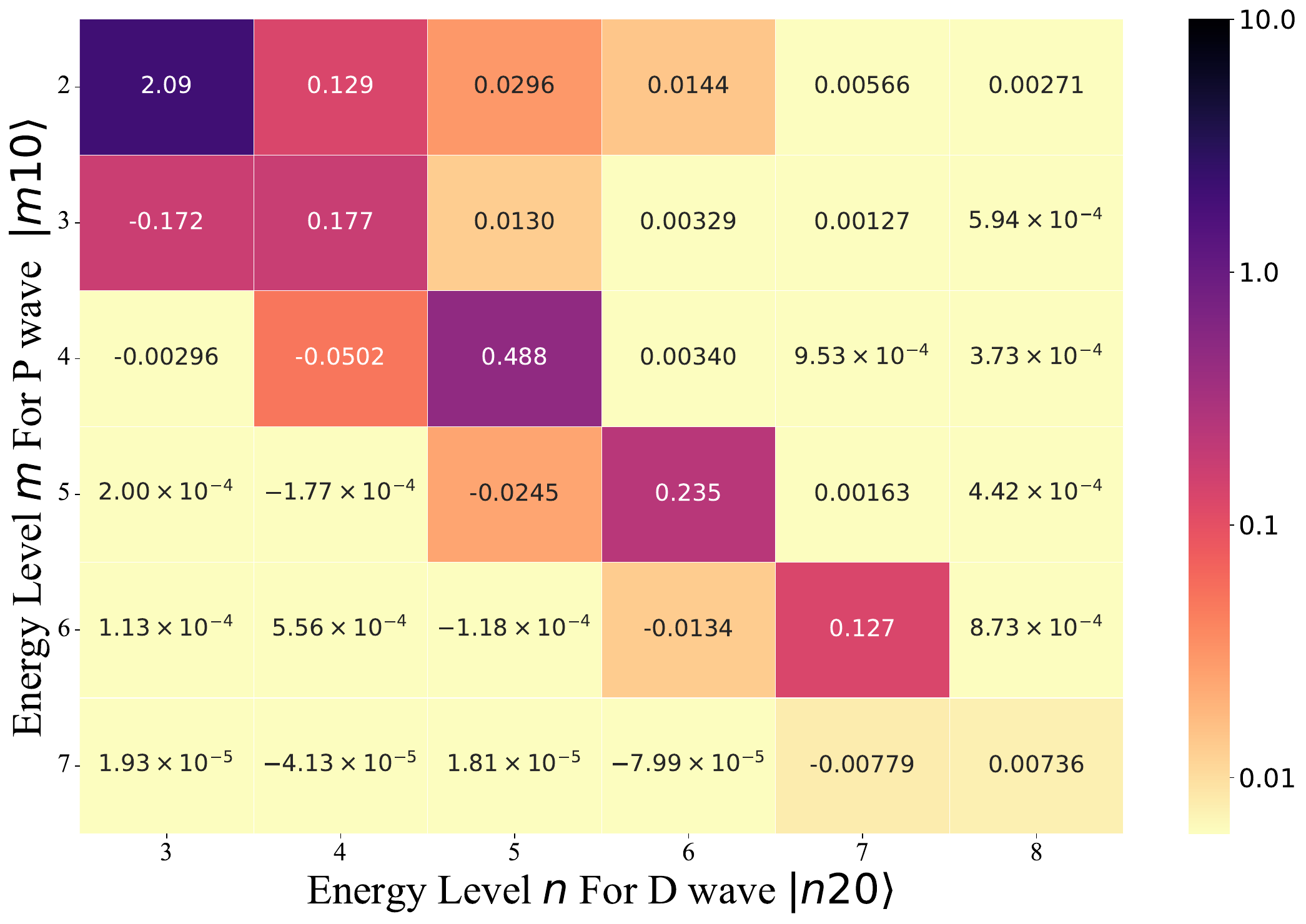}
	\caption{Heatmap for the transition coefficient $c_{nm}=\frac{\<n20|\D H|m10\>\<m10|\D H|100\>}{(E_{n20}-E_{m10})(E_{m10}-E_{100})}$ for Stark effect, all in units of $(e|\mathbf{E}|)^2/\text{GeV}^4$. The row labels the intermediate P wave state $|m10\>$. The column labels the end point D wave state $|n20\>$. The color intensity, mapped on a logarithmic scale in the color bar, corresponding to the absolute value of the coefficient. Parameters for $J/\psi$ have been used. One should see the translations to higher energy levels are excessively suppressed.} 
	\label{fig:starktranslation}
\end{figure}

We have already obtained the unperturbed eigenstates wave function. The calculation of D wave correction $\ps_D^E$ is similar to $\ps_D$ except that it involves a double sum. Again S-wave correction $\psi_S^E$ is excluded in our calculation since it does not contribute to spin alignment at leading order.
%
We have calculated first few Stark transitions from ground state S wave to P waves at $m^{th}$ level, then to D wave at $n^{th}$ level, with the corresponding magnitudes shown in FIG \ref{fig:starktranslation}. We can see that the transitions to higher levels are suppressed compared with the the nearest transition $\ket{100}\rightarrow\ket{210}\rightarrow\ket{320}$. For high accuracy, we sum up all possible translations for $m=14$ and $n=15$ in the final calculation of spin alignment.

Finally we consider Zeeman interaction, which induces mixing among triplet and singlet states. The mixing is conveniently calculated in terms of spin states defined by quantization axis $\hat{B}$. We will obtain the corresponding spin density matrix, then find the spin density matrix in spin states defined by quantization axis $\hat{l}$ through a unitary transformation. We denote the states defined by quantization $\hat{B}$ and $\hat{l}$ with primed and unprimed states respectively. In terms of the primed states, the mixing is known up to $O(B^2)$ as \cite{Alford:2013jva}
\begin{align}\label{mix_triplet}
&|10'\>_{B}=\(1-\frac{\c^2}{8}\)|10'\>+\frac{\c}{2}|00'\>,\nonumber\\
&|00'\>_{B}=\(1-\frac{\c^2}{8}\)|00'\>-\frac{\c}{2}|10'\>,
\end{align}
with $\c=\frac{2qB}{m\D E}$ and the subscript $B$ indicating states after the mixing. This leads to the spin density matrix
\begin{align}\label{rho_prime}
\r'=\text{diag}(1,1,1-\frac{\c^2}{4})
\end{align}
assuming no contribution from $|00'\>$.
Now we work out the unitary transformation between the primed and unprimed triplet states. Parameterizing the quantization axes as $\hat{l}=(0,0,1)$ and $\hat{B}=(\sin\th_b\cos\f_b,\sin\th_b\sin\f_b,\cos\th_b)$, we first find out the Euler rotations that brings $\hat{l}$ to $\hat{B}$. This can be achieved by the first two Euler rotations with precession angle $\vf=\f_b+\frac{\p}{2}$ and nutation angle $\th=\th_b$ \footnote{We adopt the $z-x-z$ convention for the Euler rotations}. The triplet states in vector basis are related by rotation as
\begin{align}\label{rotation}
\begin{pmatrix}
|V_x'\>\\
|V_y'\>\\
|V_z'\>
\end{pmatrix}=
\begin{pmatrix}
\cos\vf &\sin\vf &0\\
-\cos\th\sin\vf &\cos\th\cos\vf &\sin\th\\
\sin\th\sin\vf &-\sin\th\cos\vf &\cos\th
\end{pmatrix}
\begin{pmatrix}
|V_x\>\\
|V_y\>\\
|V_z\>
\end{pmatrix},
\end{align}
with $|V_x\>=-\frac{1}{\sqrt{2}}(|11\>-|1-1\>)$, $|V_y\>=\frac{i}{\sqrt{2}}(|11\>+|1-1\>)$, $|V_z\>=|10\>$ and similarly for primed states. Using \eqref{rotation} and the (unnormalized) primed spin density matrix \eqref{rho_prime}, we easily obtain the (unnormalized) spin density matrix for unprimed states
\begin{align}\label{sdm_Zeeman}
\r=
\begin{pmatrix}
1-\frac{\sin^2\th}{8}\c^2 &-\frac{e^{2i\vf}\sin^2\th}{8}\c^2& -\frac{ie^{i\vf}\sin(2\th)}{8\sqrt{2}}\c^2\\
-\frac{e^{-2i\vf}\sin^2\th}{8}\c^2 &1-\frac{\sin^2\th}{8}\c^2& -\frac{ie^{-i\vf}\sin(2\th)}{8\sqrt{2}}\c^2\\
\frac{ie^{-i\vf}\sin(2\th)}{8\sqrt{2}}\c^2& \frac{ie^{i\vf}\sin(2\th)}{8\sqrt{2}} \c^2& 1-\frac{\cos^2\th}{4}\c^2
\end{pmatrix}
\end{align}
As a consistency check, we confirm the corresponding photon self-energy indeed has the structure \eqref{Pi_TL}. From \eqref{Pi_rep}, we have for S-wave state that $\P^{ij}\propto\r_{AB}n_A^{*i}n_B^j$. With \eqref{sdm_Zeeman}, we readily verify that the resulting $\P^{ij}$ indeed satisfies \eqref{Pi_rep} with $\frac{\P_T-\P_L}{\P_T}=\frac{\c^2}{8}$, $\vf=\f_b+\frac{\p}{2}$ and $\th=\th_b$. It follows immediately from \eqref{alignment} that
%
\begin{align}\label{alignment_Zeeman}
\l_\th\simeq -\frac{1}{16}\(3\cos^2\th_{Bl}-1\)\c^2,
\end{align}
with $\th_{Bl}=\th$.

\section{Discussions}

\begin{figure}[h]
	\centering
	\includegraphics[width=1.0\linewidth]{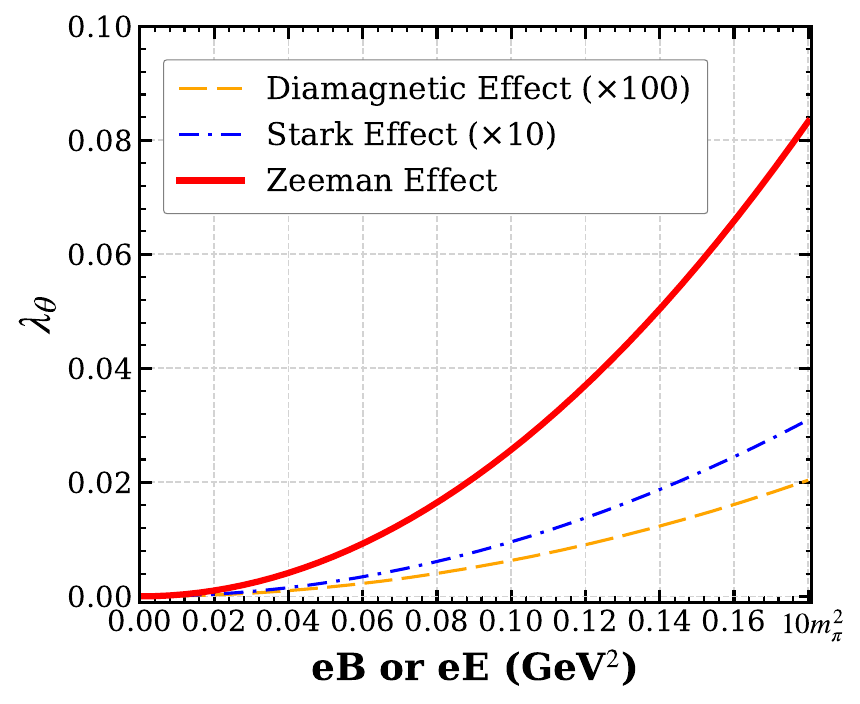}
	\caption{ Contributions to quarkonia spin alignment from different sources at their maxima reached at $\th_{Bl}=0$ or $\th_{El}=0$. Stark effect and diamagnetic effect are scaled by 10 and 100 times separately to be compared with the Zeeman effect. $m=1.84\text{GeV}$ is used corresponding to $\jps$ \cite{Eichten:1979ms}.}\label{fig:alignment}
\end{figure}

We plot the maxima of all three contributions reached at $\th_{Bl}=0$ or $\th_{El}=0$ in Fig.~\ref{fig:alignment}. In the phenomenologically motivated parameter choice made therein, a clear hierarchy of different contributions is seen, with the Zeeman effect being dominant, followed by the Stark effect and diamagnetic effect. The hierarchy can be understood from a power counting in $m$: the Zeeman effect is $O(m^{-2})$, and the Stark effect and diamagnetic effect receive suppression from $\a_k\lesssim O(m^{-2})$ in the annihilation vertex. Additionally, the two effects are suppressed by transition matrix elements and an extra factor of $O(\D E/m)$ for the diamagnetic effect. These suppression factors are partially compensated by strong EM fields: the magnetic field produced in HIC at LHC energy can reach up to $eB\simeq 10m_\p^2$ in lab frame. Boosting the magnetic field into the quarkonium frame gives both EM fields that can reach even larger magnitude. The sign and order of magnitude of the total contribution is consistent with experiments \cite{ALICE:2022dyy}. Caveats should be taken as rapid decay of the magnetic field requires further modeling. Also $\jps$ regenerated in the medium is not taken into account. Nevertheless, we expect the mechanism to give dominant contribution to spin alignment of energetic quarkonia, as they experience the early stage magnetic field and are little affected by the medium.

In order to use spin alignment as a diagnosis tool for quarkonium structure, we wish to eliminate the Zeeman contribution, which does not depend on quarkonium potential. This can be achieved by proper choice of the quantization axis. Despite the dominance of Zeeman interaction at its maximum, we can switch it off by the choice $\th_{Bl}=\cos^{-1}\frac{1}{\sqrt{3}}$ but $\th_{El}\ne\cos^{-1}\frac{1}{\sqrt{3}}$. This leaves only the Stark effect contribution, which carries valuable information on how the quarkonium shape responds to external electric field and can be a valuable probe of quarkonium structure.

\begin{figure}[h]
	\centering
	\includegraphics[width=1.0\linewidth]{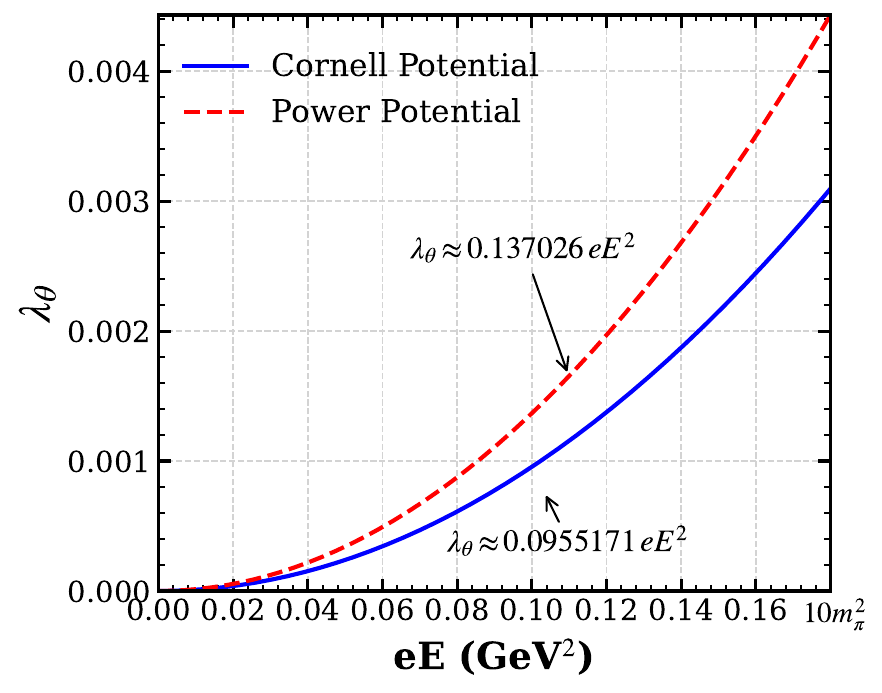}
	\caption{Comparison of spin alignment from Stark effect from Cornell potential and power potential. A significant difference of about $43\%$ is seen, as compared to less than $1\%$ difference in mass spectrum, making the spin alignment a sensitive probe of the quarkonium potential.}\label{fig:potential}
\end{figure}

%
To illustrate the sensitivity of spin alignment to structure of quarkonium, we consider prediction of spin alignment in the power potential $V=A+B r^\a$, with $A=-8.064\text{GeV}$, $B=6.8698\text{GeV}$ and $\a=0.1$ \cite{Martin:1980rm}. For the low-lying quarkonia states up to 1D states, both the Cornell potential \cite{Eichten:2002qv} and power potential \cite{Martin:1980rm} give a mass spectrum within $1\%$ of experimental values of spin-averaged charmonia states \cite{ParticleDataGroup:2020ssz}, as shown in Table \ref{mass compare}. The 1D singlet state is not yet observed experimentally \cite{Belle:2020esr}. We use $m_{\h_{c2}}-m_{\ps_{2}(1D)}=9\pm10\text{MeV}$ from the lattice study \cite{Cheung:2016bym}. Fig.~\ref{fig:potential} shows a comparison of spin alignment from Stark effect with the two potential models. A significant difference of about $43\%$ is seen, making the spin alignment a sensitive probe of the quarkonium potential. We iterate that it is achieved by proper choice of quantization axis that keeps only the Stark effect contribution to the spin alignment.
\begin{table}[h]
	\setlength{\tabcolsep}{12pt}
	\begin{tabular}{|c|c|c|c|}
		\hline
		\diagbox{State}{mass} & $\mathrm{M}^{\text{corn}}_{c\bar c}$& $\mathrm{M}^{\text{pow}}_{c\bar c}$& $\mathrm{M}^{\text{exp}}_{c\bar c}$\\
		\hline
		1S& 3.068& 3.067& 3.068\\
		1P& 3.527& 3.503& 3.525\\
		2S& 3.680& 3.670& 3.674\\
		1D& 3.817& 3.788& 3.824\\
		\hline
	\end{tabular}
	\caption{Mass spectrum for low-lying spin-averaged $c\bar c$ states, obtained separately by Cornell potential, power potential and experimental data, all in units of GeV. The mass of the unobserved $\h_{c2}$ is estimated using lattice results, see text for detailed discussions. The uncertainties in the last column are within $10^{-3}$ and not shown.}
	\label{mass compare}
\end{table}

\section{Conclusion and outlook}

We have studied the distortion of wave function for quarkonium in the presence of external electromagnetic fields. The corresponding shape change gives rise to orbital contribution of its spin alignment. We have also studied genuine spin contribution to spin alignment from spin states mixing in the presence of magnetic field. We have found the spin contribution dominates over orbital contribution, with the sign and magnitude in qualitative agreement with experiments.

We have also obtained the dependencies of each contribution on quantization axis. By a proper choice of quantization axis, it is possible to switch off the spin contribution, leaving only orbital contribution from motional Stark effect. This allows us to access the shape change of the quarkonium. We have further shown that this offers a very sensitive probe of the quarkonium potential as compared to conventional method.

Finally let us remark that the structure change induced contribution to spin alignment proposed in this work is not limited to quarkonia, but can also affect heavy mesons. Due to their short lifetimes, the mechanism is expected to play a prominent role for heavy mesons as magnetic field also peaks at early stage of HIC. Combined with recent measurement of heavy meson spin alignment \cite{ALICE:2025cdf}, the mechanism proposed in this work may also provide a useful tool for studying structure of heavy mesons.







\begin{acknowledgments}

We thank X.-Z. Bai, D.-F. Hou, Y. Jia, Z.-T. Liang, Y. Li, Y.-Q. Ma, K. Redlich, Z.-B. Tang, Q. Wang, J.-Y. Xiang, X.-B. Zhao and P.-F. Zhuang for stimulating discussions. This work is in part supported by NSFC under Grant Nos 12475148 and 12075328.
\end{acknowledgments}

\end{document}


\maketitle

In the supplementary information, we show that the current in Eq.~(19) in the main text is consistent with the NRQCD framework. To this end, we calculate the amplitude for quarkonium annihilation and recreation with both the current in Eq.~(19) and the NRQCD framework respectively. For simplicity, we consider quark carrying unit charge. In this case, the amplitude is given by
\begin{align}\label{amp_current}
i{\cal M}=(-ie)^2\<Q\bar{Q}|J^j\frac{i\d_{ij}}{Q^2}J^i|Q\bar{Q}\>,
\end{align}
where we have used the temporal axial gauge for the gauge invariant amplitude. Choosing quark/anti-quark to have 3-momentum $\pm\vp$ and $\pm\vp'$ in the initial and final quarkonia states respectively, we have from Eq.~(13)
\begin{align}\label{currents}
&J^i|Q\bar{Q}\>=\cnr^\dagger\(\s^i-\frac{1}{E_p(m+E_p)}p^ip^k\s^k\)\fnr,\nonumber\\
&\<Q\bar{Q}|J^j=\fnr^\dagger\(\s^j-\frac{1}{E_p(m+E_p)}p'{}^jp'{}^l\s^l\)\cnr.
\end{align}
By conservation of energy, we have $p^2=p'{}^2$ and $Q^2=m^2+p^2$. Plugging \eqref{currents} into \eqref{amp_current} and expand to $O(\frac{p^2}{m^2})$, we obtain
\begin{align}\label{amp_vertex}
{\cal M}=-\frac{\p\a}{m^2}\big[(1-\frac{p^2}{m^2})\cnr^\dagger\s^i\fnr\fnr^\dagger\s^i\cnr-\frac{1}{2m^2}(p^ip^j+p'{}^ip'{}^j)\cnr^\dagger\s^i\fnr\fnr^\dagger\s^j\cnr\big].
\end{align}
Now we obtain the same amplitude in the NRQCD framework. While the explicit coefficients are not available for electromagnetic annihilation vertices, they can be obtained from color octet strong annihilation vertices by the replacement $T^a\to1$, because analogous diagram contributes to the electromagnetic annihilation process. At lowest order in $\a$, only S-wave and D-wave coefficients are nonvanishing. Adding up (A5) and (A7) of \cite{Bodwin:1994jh} and reinstate external states fields, we obtain
\begin{align}\label{amp_NRQCD}
&{\cal M}=\frac{f_8({}^3S_1)}{m^2}\cnr^\dagger\s^i\fnr\fnr^\dagger\s^i\cnr+\frac{3g_8({}^3S_1)-g_8({}^3S_1,{}^3D_1)}{3m^2}v^2\cnr^\dagger\s^i\fnr\fnr^\dagger\s^i\cnr\nonumber\\
&+\frac{g_8({}^3S_1,{}^3D_1)}{2m^2}(v^iv^j+v'{}^iv'{}^j)\cnr^\dagger\s^i\fnr\fnr^\dagger\s^j\cnr.
\end{align}
Using (A6) and (A8) of \cite{Bodwin:1994jh} with the replacement $\a_s\to\a$, we have
\begin{align}
f_8({}^3S_1)\to-\p\a,\quad g_8({}^3S_1)\to\frac{4\p}{3}\a,\quad g_8({}^3S_1,{}^3D_1)\to\p\a.
\end{align}
Plugging the above into \eqref{amp_NRQCD} and noting at this order $v^i\simeq \frac{p^i}{m}$, $v'{}^i\simeq \frac{p'{}^i}{m}$, we readily confirm the agreement with \eqref{amp_vertex}.